\documentclass[aps,prd,article]{revtex4}
\usepackage{amsmath}% http://ctan.org/pkg/amsmath
\begin{document}

\title{Quantum field theory treatment of magnetic effects on a system of free electrons}
\author{C. Verzegnassi}
\email{claudio@ts.infn.it }
\affiliation{Politecnico di Ingegneria e Architettura, University of Udine, Udine, Italy, and AMeC (Association for Medicine and Complexity), Trieste, Italy}
\author{R. Germano}
\affiliation{PROMETE CNR Spin Off, Napoli, Italy}
\author{P. Kurian}
\email{pkurian@howard.edu}
\affiliation{National Human Genome Center and Department of Medicine, Howard University College of Medicine, Washington DC, USA; Computational Physics Laboratory, Howard University, Washington DC, USA}

\begin{abstract}
$$ $$
The effects of a magnetic field on the energy and on the spin of free electrons are computed in the  framework of quantum field theory. In the case of a constant moderate field and with relatively slow electrons, the derived formulae are particularly simple. A comparison with the approaches of classical physics and of quantum mechanics shows essential differences and important analogies. The relevance to the magnetic effects of the initial polarization components of the electron states and the possible existence of special values of these quantities are discussed in the final conclusions, which might be useful to explain recent experiments on quasi-free electrons in chiral systems in biology.
%The effects of a magnetic field on the energy and on the spin of free electrons are computed in the theoretical framework of Quantum Field Theory. In the case of a constant field and of relatively slow electrons the derived formulae are particularly simple. A comparison with the approaches of Classical Physics and of Quantum Mechanics shows essential differences and important analogies. The relevance on the effects of the initial  spin  components of the electron and the existence of special optimal values for them are discussed in the final conclusions. 
\end{abstract}

\maketitle

%\endfrontmatter
 
 %The idea that a magnetic field might produce positive effects on the human organism has produced  strong and motivated efforts in recent times (1). Several different applications have been performed, and medical and biological detailed illustrations of the obtained results have been provided (2) that might appear, least to say, encouraging . In this spirit, one might try to produce a scientific theoretical prediction of the observed effects. At the moment, such a prediction does not exist. Certainly, one of the reasons for this lack is the fact that the magnetic field should interact with the crowd of electrons and atomic nuclei that the organism contains.
Two of us have described in a recent letter \cite{PLA2016} the reciprocal and compensatory conversion between the spin and orbital angular momentum of a free electron induced by a magnetic field in the theoretical framework of quantum electrodynamics. To describe a more realistic system, with a crowd of electrons and atomic nuclei in full generality, the elementary properties of these constituents should determine the results of the process of interaction with the magnetic field, even if emergent or novel effects arise at each scale of organization in the system. However, the role of the mutual interactions between electrons and nuclei will make the calculation of the purely magnetic effects hardly performable. An almost obvious statement is that the suggested calculation might be more easily carried out if the mutual interaction between these constituents were extremely weak, such that the interaction might be ignored altogether. The simplest scenario would be the case of a system of free electrons, of which there are abundant examples in solid-state, condensed matter, and biological physics.

%The main question, before performing this simplified calculation, would be the existence of a  possible relevance of this search for the human organism. In other words, is there a possible relevant role played by systems of free electrons at least in certain components  of the organism? The answer to this question is almost completely positive. It is provided by the rigorous theoretical study that has led to the separation of the organic water in two components, one of which, usually called "separate water" is essentially made of "nearly free" electrons, and for a detailed discussion of these points we defer to the existing literature (3). 

In this spirit we have considered the goal of computing the effects of a magnetic field on a system of completely free electrons, considering this calculation as an initial step to which suitably weak interactions can be added later. %To compute magnetic effects on systems of free electrons, a very reasonable starting point appears to us that of first computing these effects on the simplest case of one single free  electron. This calculation has  actually already been performed, and the results are published in two separate papers, the first of which derives the effect on the electron spin (4), whilst the second one derives the effect on the electron energy (5). The theoretical framework that has been chosen is that of Relativistic Quantum Field theory , and we shall devote a short comment to illustrate the reasons of this choice . 
To summarize these effects, we shall now review the main expressions for a single electron state that are required to obtain the principal results for a system of free electrons. In our work, we have followed the quantum field theory notations and conventions of Peskin and Schroeder \cite{Peskin}.    

For a single electron state we shall choose as basic quantities the four complex components of the electron field $\psi(\mathbf{x})$, written as $\psi_s(t,\vec{x})$, where $s=1,2,3,4$. Using the explicitly hermitian form of the Dirac Lagrangian, we have rewritten below the total energy in terms of these components:
\begin{eqnarray}
\mathcal{H} = \int d^3x && \left[ -\frac{i}{2}\overline{\psi} \gamma^{k} \partial_{k} + \frac{i}{2} (\partial_{k}\overline{\psi})\gamma^{k} + m_e \overline{\psi} \right] \psi \nonumber\\
= \int d^3x && \left[2m_e \, \text{Re}\left(\, \psi_1^\ast \psi_3 + \psi_2^\ast \psi_4\,\right) \right. - \left.  \,\text{Im}\left(\, \psi_1^\ast \partial_3 \psi_1 +\psi_1^\ast \left(\, \partial_1 -i\partial_2\,\right)\psi_2 \right.\right. +\left.\left. \psi_2^\ast \left(\, \partial_1 +i\partial_2\,\right)\psi_1 -\psi_2^\ast \partial_3 \psi_2  \,\right)\right.+  \nonumber\\
&&\left. \,\text{Im}\left(\, \psi_3^\ast \partial_3 \psi_3 +\psi_3^\ast \left(\, \partial_1 -i\partial_2\,\right)\psi_4\right. \right. + \left.\left. \psi_4^\ast \left(\, \partial_1 +i\partial_2\,\right)\psi_3 -\psi_4^\ast \partial_3 \psi_4  \,\right)
\right.].
\label{ham}
\end{eqnarray}
Consistent with the defined quantities and notations for the changes in the electron field components with the introduction of purely magnetic potentials $\vec{A}$, we also derived the effects of these potentials on spin \cite{PLA2016} and energy \cite{ClaudioJMP}. In the non-relativistic limit (NRL) where we are concerned ($\psi_1, \psi_2 \gg \psi_3, \psi_4$), 
\begin{equation}
\Delta _{\vec{A}} \mathcal{H} \xrightarrow{\text{NRL}} -4\vert e\vert \int d^3x \, \vec{A}\cdot \vec{s}^{\,\prime}
\label{nrl}
\end{equation}
where $e$ is the electron charge and $\vec{s}^{\,\prime}$ is the ``spin current'' in the NRL. The spin current is defined generally as the quantity whose integral gives the spin vector itself, $\vec{S}=\int d^3x \, \vec{s}$, and its components are 
\begin{eqnarray}
\nonumber s_1=&& \text{Re}\left(\,\psi_1^\ast \psi_2 + \psi_3^\ast \psi_4\, \right)\label{s1}\\ 
\nonumber s_2=&& \text{Im}\left(\,\psi_1^\ast \psi_2 + \psi_3^\ast \psi_4\, \right)\label{s2}\\ 
s_3=&& \frac{1}{2}\left(\,\psi_1^\ast \psi_1 - \psi_2^\ast \psi_2\right. + \left. \psi_3^\ast \psi_3 - \psi_4^\ast \psi_4 \,\right).
\label{s3}
\end{eqnarray}

%\begin{eqnarray}
%&& \Delta_A \psi_1 =\frac{|e|}{m} \left(\, A_0 \psi_3 + A_1 \psi_4 -i A_2 \psi_4 +A_3 \psi_3\,\right)\ ,\nonumber\\
%&& \label{delta1}\\
%&& \Delta_A \psi_2 =\frac{|e|}{m} \left[\, A_0 \psi_4 + A_1 \psi_3 +i A_2 \psi_3 -A_3 \psi_4\,\right]\ ,\nonumber\\
%\label{delta2}\\
%&& \Delta_A \psi_3 =\frac{|e|}{m} \left[\, A_0 \psi_1 - A_1 \psi_2 +i A_2 \psi_2 -A_3 \psi_1\,\right]\ ,\nonumber\\
%&& \label{delta3}\\
%&& \Delta_A \psi_4 =\frac{|e|}{m} \left[\, A_0 \psi_2 - A_1 \psi_1 -i A_2 \psi_1 +A_3 \psi_2\,\right]\ .\nonumber\\
%&& \label{delta4}
%\end{eqnarray}

%\begin{equation}
%\vec{s}_{(NRL)}=\left[\, Re\left(\, \psi^\ast_1 \psi_2\,\right)\ , Im\left(\, \psi^\ast_1 \psi_2 \,\right)\ ,
   % \frac{1}{2}\left(\, \psi^\ast_1 \psi_1 - \psi^\ast_2 \psi_2  \,\right)\, \right]
%\label{nrl4}
%\end{equation}

A short discussion of Equation (\ref{nrl}) will be relevant here. Consider the expression that would have been predicted for the magnetic effect on the energy by a strictly quantum mechanical approach. As is well known, this expression---called the Zeeman effect---can be written for an electron as a scalar 
product between the electron angular momentum $\vec{J}$ and the magnetic field $\vec{B}$. For the contribution coming from the electron spin $\vec{S}$, we would have
\begin{equation}
\Delta _{\vec{B}} \mathcal{H} = -g_s \mu_B \vec{B} \cdot \vec{S}
\label{Zeeman}
\end{equation}
where $g_s \approx 2$ is the gyromagnetic factor and $\mu_B=|e|/2m_e$ is the Bohr magneton. One can see a faint resemblance between Equation (\ref{Zeeman}) above and the quantum field theory expression of Equation (\ref{nrl}), with (roughly speaking) a replacement of the magnetic field $\vec{B}$ with the magnetic potentials $\vec{A}$ and likewise of the spin $\vec{S}$ with the spin current $\vec{s}^{\,\prime}$. %In other words, to compute the effect on a free electron, one cannot simply use the QM expression (13) but must move to the Quantum Field theory calculation.
Using the same quantum field theory framework, we have previously derived the effect of a magnetic field on the electron spin, which in full generality is given by the following expression: 
%(15) EQ.(9) OF REF. (4)
\begin{equation}
\Delta_{\vec{A}}\vec{S}=\vert e \vert \int d\vec{x}\left[\, \vec{A}\times \vec{\rho}_E\,\right],
\label{DeltaS}
\end{equation}
where  $\vec{\rho}_E$ is defined as
%(16) EQ. (14) OF (Extra)
\begin{equation}
\vec{\rho}_E=\frac{i}{m_e}\psi^\dagger \vec{\gamma} \psi 
%&& = \frac{2}{m}\left(\, Im\left(\psi_4^\ast \psi_1  + \psi_3^\ast \psi_2\,\right)\ , Re\left(\, \psi_4^\ast \psi_1 - \psi_3^\ast \psi_2\,\right)\ ,
%-Im\left(\psi_1^\ast \psi_3  + \psi_4^\ast \psi_2\,\right)\,\right]
\end{equation}
with $\vec{\gamma}$ including the conventional gamma matrices. 
%Starting from Eq.(14) and looking for more relevant conclusions, we have computed the magnetic effect on the third component of the Spin:
%(17)EQ. (12) OF REF.(4)
%\begin{equation}
% \Delta_{\vec{A}}S_3=\vert e \vert \int d\vec{x}\left[\, A_1 \rho_{E\, 2} - A_2 \rho_{E\,1}\,\right]
%\end{equation}
%We have then considered the realistic case of a constant magnetic field B which we choose to oriented along the z axis,
% $\vec{B}=(0\ ,0\ , B_3)$. The expressions of the magnetic potential are then fixed by:
%(18) EQ. (13) OF REF. (4) WITH H REPLACED BY B
%\begin{equation}
 %\vec{A}(\vec{x})=\frac{1}{2}\vec{B}\times \vec{x}
%\end{equation}
%which leads to the expression:
%(19) EQ. (14) OF REF.(4)  with H3 replaced by B3
%\begin{equation}
 %\Delta_{\vec{A}}S_3=\vert e \vert \frac{B_3}{2}\int d\vec{x}\left[\, x_2 \rho_{E\, 2} + x_1 \rho_{E\,1}\,\right]
%\end{equation}
Such an expression in Equation (\ref{DeltaS}) is only possible in quantum field theory, where the value for the spin in terms of the fields is known (see Equations (\ref{s3})). 

We then computed the expectation value of the change in spin for a single electron state. More precisely, we have considered a properly normalized free electron state of momentum $\vec{k}=(0,0,k_3)$ along the $z$ axis, in a linear superposition of spin eigenstates: 
\begin{equation}
\left |\Psi(\vec{k}) \right \rangle = \lambda_+ \left|\uparrow, \vec{k} \right \rangle + \lambda_- \left|\downarrow,\vec{k}\right \rangle.
\end{equation}
Under the assumption that the components of the magnetic potential remain reasonably close to their average values in the finite volume of dimension $d$ where the applied magnetic potential is non-vanishing, we obtained the following result for the expectation value of the change for the third component of spin:
\begin{equation}
\left \langle\Delta_{\vec{A}} S_3 \right \rangle= \frac{|e|}{m_e} B_3 d \left[\, \text{Re}(\, \lambda_+\lambda_-^\ast\,) + \text{Im}(\, \lambda_+\lambda_-^\ast)\,\right],
\label{DeltaS3}
\end{equation} 
where $\lambda_+, \lambda_-$ are the complex coefficients defining a linear superposition of spin eigenstates.
%More precisely, we have considered a free electron state of momentum k=(o,o, k3) along the z axis i.e. parallel to the magnetic field. We have parametrized this state as a linear superposition of two spin eigenstates:
%(20)EQ. (17) OF REF. (4)
%\begin{equation}
 %\vert \psi(\vec{k})>=\lambda_+ \vert \uparrow\ ,\vec{k}> +\lambda_- \vert \downarrow\ ,\vec{k}>
%\end{equation}
%with
%\begin{equation}
 %|\lambda_+|^2 + |\lambda_-|^2 =1.
%\end{equation}
 
There are two features of Equation (\ref{DeltaS3}) that should be highlighted. First, in classical physics there exists a known effect that a static magnetic field $B$ has on the motion of 
a magnetic moment. This motion is characterized by the well-known €œLarmor precession, where the magnetic moment rotates about the magnetic field vector, describing  a cone around the axis of the applied field. In the rotation, the Larmor angular velocity $\omega_L$ has the value
\begin{equation}
\omega_L= g_s \, \mu_B \, B.
\label{Larmor}
\end{equation}    
Without loss of generality, the magnetic field can be oriented along the $z$ axis and therefore in our discussion $B$ can be equated with $B_3$. The circular motion of the magnetic moment on a circumference of radius $r$ perpendicular to the field direction has a velocity $v$ given by the usual expression for centripetal motion:
\begin{equation}
 v = r \omega_L. 
\end{equation}
During the motion, the magnetic moment acquires an orbital angular momentum whose component $L_{z}$ in the direction of the field is given by the classical formula 
\begin{equation}
L_{z}= M v r = \left(M r^2 \right) \omega_L.
\label{classicalL}
\end{equation}
The acquired angular momentum component $L_z$ is therefore proportional to the Larmor frequency $\omega_L$.

If the considered system is a particle with spin angular momentum, it would be reasonable to suppose that a similar effect might appear for its spin component along the field, since its interaction with the field is analogous to that of a magnetic moment. In fact, this effect in the quantum field theory
description is given by Equation (\ref{DeltaS3}), which can be rewritten as
\begin{equation}
\left \langle \Delta_{\vec{A}} S_3 \right \rangle =(g_s \mu_B B_3) d \left[\, \text{Re}(\lambda_+\lambda_-^\ast)   + \text{Im}(\lambda_+\lambda_-^\ast)\,\right]  
=\omega_L d \, f(\lambda_+, \lambda_-).
\label{LarmorS3} 
\end{equation}
As possibly expected, the resulting spin effect in Equation (\ref{LarmorS3}) is actually proportional to $\omega_L$, which is certainly analogous to the classical result for a magnetic moment described in Equation (\ref{classicalL}). Note well that in the equation above we have defined the special function $f(\lambda_+, \lambda_-)$, which will become relevant in the discussion that follows.

The second feature of Equation (\ref{DeltaS3}) concerns the identification of the quantities that
produce the magnetic effect. One of these is, as one would expect, the magnetic field strength $B_3$. 
For the electron contribution, besides the charge-to-mass ratio, one can see that it is produced by a special function $f$ of the electron spin-up and spin-down polarization components, $\lambda_+$ and $\lambda_-$. Given the fact that $f$ determines the observable effect, it seems natural to us to associate to this factor a specific name that is descriptive of the phenomenon. One can express these features by saying that $f$  behaves like a ``chirality index'' (CI). In this spirit, we shall identify it with this œCI€ notation and say that the average electron $S_z$ magnetic variation is proportional to the electron CI.   

As a first simplified case, we consider a situation in which both components  $\lambda_+$ and $\lambda_-$ are real. The surviving effect in Equation (\ref{DeltaS3}) is then proportional to the product $\lambda_+ \lambda_-$. This simplified CI has the following special features: (A) It vanishes when the electron state is fully polarized (maximum chirality), either in the extreme right-handed state ($\lambda_+=1$) or in the extreme left-handed state ($\lambda_-=1$). (B) It has a maximum ($1/2$) in the most symmetrical situations where $\lambda_+$ and $\lambda_-$ are equal (no chirality), with modulus $1/\sqrt{2}$, as imposed by the completeness relation for the superposed spin state. (C) It has a minimum ($-1/2$) in the most œanti-symmetrical situations where $\lambda_+$ and $\lambda_-$ are opposite (also no chirality), again with modulus $1/\sqrt{2}$. The features (A), (B), and (C) thus make the connection between the CI and the values of the probability amplitudes for the electron spin state.
 
A related question that arises at this point is the determination of the average magnetic variation of the electron energy $\mathcal{H}$. Assuming identical properties of the magnetic field previously considered and taking the same kind of electron state, with $E_0 \gg m_e$ (but still in the NRL), one finds for the average energy magnetic variation the following expression:
%(27)Eq. 14 of ref 5, DeltaH = |e| 2 lambda+lambda- Hz d2
\begin{equation}
\left \langle \Delta_{\vec{A}} \mathcal{H} \right \rangle = \frac{2 m_e |e|}{E_0} B_3 d \left[\, \text{Re}(\lambda_+\lambda_-^\ast)   + \text{Im}(\lambda_+\lambda_-^\ast)\,\right],
%2\vert e\vert \lambda_+\lambda_- \left[\,H_3 d_2- H_2 d_3 \,\right] 
 \label{DeltaH}
\end{equation}
where, again, $d$ is the dimension of the volume where the magnetic field and the electron interact non-negligibly. The main conclusion to be gleaned from Equation (\ref{DeltaH}) is that the electron CI also determines the result for the average magnetic variation of the electron energy, as it did with the spin angular momentum. We can therefore conclude that the electron CI is an essential electron quantity in the quantum field theory determination of the magnetic effects on a free electron state.

It remains an open question whether for a free electron state (or a system of free electrons) the value of its CI may be arbitrary or rather fixed by a general rule. %We have tried to answer this question proposing a personal solution that appears to us reasonably motivated, and is based on the arguments that follow.We have considered the interaction of a free electron with a magnetic field and derived in a simplified  case Eq. (27). 
Equation (\ref{DeltaH}) gives the energy that the electron will absorb from the applied magnetic field, and given the fact that the electron is almost free, it appears to us reasonably motivated that its absorption from the magnetic field should be maximal. If this assumption is accepted, the immediate consequence is that $\left|\text{Re}(\lambda_+\lambda_-^\ast)   + \text{Im}(\lambda_+\lambda_-^\ast)\right|$ attains its maximum value of $1/2$, or in the simplified case of real coefficients,
\begin{equation}
 \lambda_+= \lambda_- = \pm 1/\sqrt{2}.
 \label{reallambdas}
\end{equation}
Equation (\ref{reallambdas}) is valid for a single electron state. In our opinion, if one considers a system of completely free electrons that interact with a magnetic field, the requirement of maximum energy absorption should continue to be valid for each separate electron, since it does not interact even weakly with the other components of the system. Thus for a system of free electrons, each CI of the various electron components should retain the value of Equation (\ref{reallambdas}). Alternatively stated, for each electron of the system, the probability that its spin component is found in the direction of the field will be equal to the probability that it is found in the opposite direction. This conclusion is indeed satisfying since it is in total agreement with the results of the historical Stern-Gerlach experiment.%notwithstanding the difficulties of observing this effect with free electrons due to the large deflection in trajectory of charged particles moving through a magnetic field.

In a strictly quantum mechanical description of a system of free electrons, the system acquires from its interaction with a magnetic field $B$ a quantity called magnetic susceptibility due to the fact that the number of electrons with spin in the direction of the field is different from the number of electrons with spin in the opposite direction. The spin probability amplitude is not changed by the introduction of a magnetic field. On the contrary, one can conclude from Equation (\ref{DeltaS3}) that the average value of the increase of the electron spin component in the direction of the magnetic field is proportional to the magnetic field component, and this value can be larger or smaller than the spin component in the opposite direction, depending on the values of the spin probabilities that determine the state polarization. This represents a quantum field theory generalization of the quantum mechanics result.         

Clearly, in a biological situation the states to be considered are much more complicated than free electron systems with specific CIs. However, the role of spin in biology may still be ubiquitous, as recent studies have confirmed the spin selectivity of electron transport through chiral molecules \cite{Naaman, Naaman2} such as DNA and proteins. Furthermore, recent experimental findings by one of us obtained in V. Elia's research group \cite{Germano1} show that supramolecular structures can organize in bi-distilled liquid water due to low-energy physical perturbations of various kinds. %These large nanostructures are formed by aggregates of water molecules, organized coherently according to water's unique electronic structure, and behave quite intriguingly like biological matter. First, they demonstrate chirality as determined by standard circular dichroism measurements, and second, they exhibit ultraviolet fluorescence, as evidenced by the typical $n \rightarrow \pi^*$ electronic transition. 
With the addition of infrared energy, these water nanostructures have also been shown capable of producing and conducting a significant flow of quasi-free electrons without the presence of exogenous ions \cite{Germano2}. 

The presence of a surplus of quasi-free electrons and the formation of coherent domains in liquid water---not predicted by standard views---may be described via quantum field theory \cite{Vitiello, Kurian}. Our calculations above suggest modifications to the strictly quantum mechanical calculation of the Zeeman effect on a free electron state, thus obtaining a formal analogy with the Larmor precession formula. This result may explain the formation of coherent vortices in ensembles of free electrons in liquid water \cite{Vitiello2}.

The most immediate possible extrapolation of our presented estimates would be to such large systems of free, or quasi-free, electrons. Such a realistic extrapolation of the results of our preliminary findings would require very strong theoretical and experimental efforts, which would be novel in their approach. These efforts have already begun and involve a group of dedicated colleagues, including the present authors.

%Our final conclusion is that the derivation of the various magnetic effects on free electron states offered by Relativistic Quantum Field theory appears to be original and potentially relevant. The next step appears to us to be that of understanding the possible relevance of our results for the human organism case. This requires a study of the potential role of systems of free electrons inside the organism,  that has been already emphasized in the existing literature (3). We believe that a basis of our program could be represented by the content of this paper, and starting from this  idea we have initiated our search, hoping that positive results will appear in a not too far future.
%\begin{acknowledgments}
%PK would like to acknowledge the support of the U.S. -- Italy Fulbright Commission and the Whole Genome Science Foundation. These organizations had no involvement in the study design or the presentation of results in this letter.
%\end{acknowledgments}


\begin{thebibliography}{99}
\bibitem{PLA2016}
P. Kurian and C. Verzegnassi. Phys. Lett. A \textbf{380,} 394-396 (2016).
\bibitem{Peskin}
M. E. Peskin and D. V. Schroeder. \textit{An Introduction to Quantum Field Theory}. Perseus, 1995.
\bibitem{ClaudioJMP}
C. Verzegnassi. J. Mod. Phys. \textbf{6,} 88-93 (2015). 
\bibitem{Naaman}
V. Kiran, S. R. Cohen, and R. Naaman. J. Chem. Phys. \textbf{146,} 092302 (2017).
\bibitem{Naaman2}
D. Mishra, T. Z. Markus, R. Naaman, M. Kettner, B. G\"{o}hler, H. Zacharias, N. Friedman, M. Sheves, and C. Fontanesi. Proc. Natl. Acad. Sci. USA \textbf{110,} 14872-14876 (2013).
\bibitem{Germano1}
V. Elia, T.A. Yinnon, R. Oliva, E. Napoli, R. Germano, F. Bobba, and A. Amoresano. Water \textbf{8,} 1-29 (2017).
\bibitem{Germano2}
R. Germano, E. Del Giudice, A. De Ninno, V. Elia, C. Hison, E. Napoli, V. Tontodonato, F. P. Tuccinardi, and G. Vitiello. Key Engineering Materials \textbf{543,} 455-459 (2013).  
\bibitem{Vitiello}
E. Del Giudice and G. Vitiello. Phys. Rev. A \textbf{74,} 022105 (2006).
\bibitem{Kurian}
P. Kurian, A. Capolupo, T. J. A. Craddock, and G. Vitiello. arXiv:1608.08097 [physics.bio-ph] (2017). 
%Submitted to Phys. Lett. A on 31 March.
\bibitem{Vitiello2}
E. Del Giudice and G. Vitiello. Water \textbf{2,} 133-141 (2011).
\end{thebibliography}
\end{document}